\documentclass[sigconf]{acmart}
\usepackage{amsmath}
\usepackage{graphicx}
\usepackage{hyperref}
\usepackage{subcaption}
\usepackage[linesnumbered, noresetcount, ruled,vlined]{algorithm2e}
\usepackage{algorithmic}
\usepackage{graphicx}
\usepackage{multirow, makecell}
\usepackage{array}

\AtBeginDocument{%
  \providecommand\BibTeX{{%
    \normalfont B\kern-0.5em{\scshape i\kern-0.25em b}\kern-0.8em\TeX}}}

\copyrightyear{2022}
\acmYear{2022}
\setcopyright{acmcopyright}\acmConference[CODS-COMAD 2022]{5th Joint International Conference on Data Science & Management of Data (9th ACM IKDD CODS and 27th COMAD)}{January 8--10, 2022}{Bangalore, India}
\acmBooktitle{5th Joint International Conference on Data Science \& Management of Data (9th ACM IKDD CODS and 27th COMAD) (CODS-COMAD 2022), January 8--10, 2022, Bangalore, India}
\acmPrice{15.00}
\acmDOI{10.1145/3493700.3493713}
\acmISBN{978-1-4503-8582-4/22/01}

\begin{document}

\title{Extraction of Product Specifications from the Web - Going Beyond Tables and Lists}


\author{Govind Krishnan Gangadhar}
\affiliation{%
  \institution{Clustr}
  \city{Bangalore}
  \country{India}
  }
\email{govindg0204@gmail.com}

\author{Ashish Kulkarni}
\affiliation{%
  \institution{Clustr}
  \city{Bangalore}
  \country{India}
  }
\email{kulashish@gmail.com}

\renewcommand{\shortauthors}{Gangadhar and Kulkarni}

\begin{abstract}
E-commerce product pages on the web often present product specification data in structured tabular blocks. Extraction of these product attribute-value specifications has benefited applications like product catalogue curation, search, question answering, and others. However, across different Websites, there is a wide variety of HTML elements (like <table>, <ul>, <div>, <span>, <dl>, {\em etc.}) typically used to render these blocks that makes their automatic extraction a challenge. Most of the current research has focused on extracting product specifications from tables and lists and, therefore, suffers from recall when applied to a large-scale extraction setting. In this paper, we present a product specification extraction approach that goes beyond tables or lists and generalizes across the diverse HTML elements used for rendering specification blocks. Using a combination of hand-coded features and deep learned spatial and token features, we first identify the specification blocks on a product page. We then extract the product attribute-value pairs from these blocks following an approach inspired by wrapper induction. We created a labeled dataset of product specifications extracted from 14,111 diverse specification blocks taken from a range of different product websites. Our experiments show the efficacy of our approach compared to the current specification extraction models and support our claim about its application to large-scale product specification extraction.
\end{abstract}

\begin{CCSXML}
<ccs2012>
   <concept>
       <concept_id>10002951.10003317.10003347.10003352</concept_id>
       <concept_desc>Information systems~Information extraction</concept_desc>
       <concept_significance>500</concept_significance>
       </concept>
 </ccs2012>
\end{CCSXML}

\ccsdesc[500]{Information systems~Information extraction}

\keywords{Information Retrieval, Data Mining, Extraction, Text Classification}



\maketitle

\section{Introduction}
\begin{figure}[htp]
    \centering
    \includegraphics[width=.85\columnwidth]{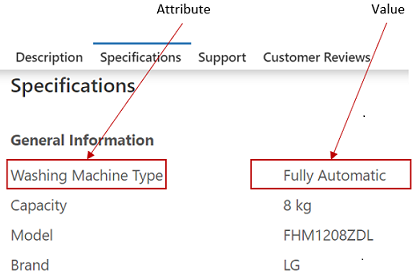}
    \caption{Specification block for a washing machine product}
    \label{fig:spec_example}
\end{figure}

There has been a massive spurt of information on the World Wide Web in the past decade, with one of the primary drivers of this growth being the arrival of e-commerce websites. These websites are a source of large-scale product data, often to the tune of hundreds of thousands to millions. The information about a product is typically presented on a webpage in a combination of structured and unstructured formats. We are specifically interested in the product specifications available as attribute-value pairs in structured tabular blocks.
%The product web pages have a diverse amount of information represented in the form of HTML, with the crucial information that describes the product concentrated as specification. This specification information of a product alone is enough to describe everything about it. 
Figure~\ref{fig:spec_example} illustrates the specifications for a washing machine product taken from an online e-commerce website.  %presented under multiple sections which we call as specification blocks. 
We refer to this as a \textit{specification block}. Customers often use the information in these specification blocks to compare products and make their purchase decisions. %This is used as one of the quickest ways to compare one product against another and is often the only source of information that people take into account when it comes to making any product related decision. Specification information is often represented in a condensed format as attribute-value pairs which define the properties of the products in a condensed form. 
The product specification data has also been shown to benefit applications like faceted search~\cite{DBLP:conf/semweb/StolzH15a}, question answering~\cite{10.1145/3308560.3316597,10.1145/3289600.3290992}, product ontology curation~\cite{LEE200616}, canonicalization of product attributes, product comparison~\cite{gopalakrishnan2012matching}, offer-product matching~\cite{10.1145/2020408.2020474} and creating structured product catalogs ~\cite{DBLP:journals/corr/abs-1105-4251}. Thus, automatic extraction of product specifications at scale is an important research problem with practical interest.

\begin{figure*}[htp]
%\begin{subfigure}{.5\textwidth}
\begin{subfigure}{\columnwidth}
  \centering
  \includegraphics[width=.9\columnwidth, height=5.3cm]{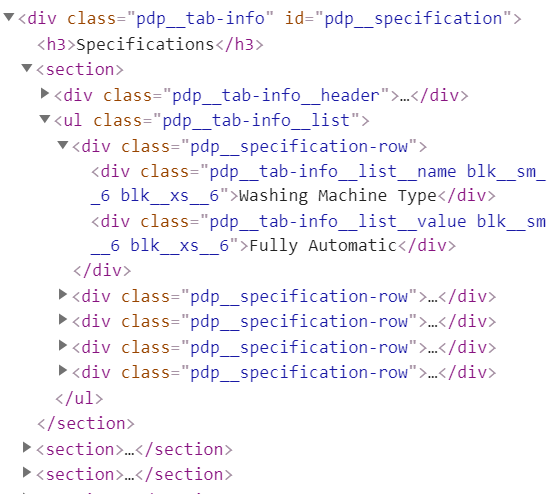}
  \caption{A specification block structured using <ul>-<div>-<div> tags }
  \label{fig:sfig1}
\end{subfigure}%
%\begin{subfigure}{.5\textwidth}
\begin{subfigure}{\columnwidth}
  \centering
  \includegraphics[width=.9\columnwidth, height=5.3cm]{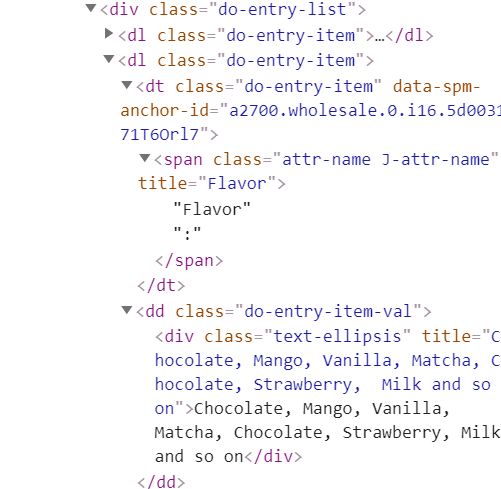}
  \caption{A specification block structured using <dl>-<dt>-<span> tags}
  \label{fig:sfig2}
\end{subfigure}
\caption{Specification blocks taken from two different product websites illustrating the heterogeneity of HTML tags used for their rendering}
\label{fig:inspect_elements_example}
\end{figure*}

Although specification blocks typically appear in a tabular structure, different e-commerce websites use a wide range of HTML tags to render them, with tags, such as, <table>, <li>, <div>, <dl>, <dt>, <dd> {\em etc.} being a common occurrence. In fact, such heterogeneity in specification blocks is seen across product pages within a product website as well. Refer to Figure~\ref{fig:inspect_elements_example} that illustrates this heterogeneity of specification blocks for two product pages taken from two popular e-commerce websites. This poses a challenge in the automatic extraction of specifications from these blocks. Previous works have focused on extracting information from specification blocks composed of only table tags~\cite{10.1145/1772690.1772814,10.1145/1242572.1242583, 10.3115/990820.990845}, with the recent ones~\cite{qiu2015dexter, 10.1145/3106426.3106449} adding capability for lists as well. Compared to the other HTML tags used for structuring specification blocks, tables and lists are relatively easier to interpret and have a clear notion of rows and columns. The aforementioned approaches leverage these signals in several hand-coded features to train classifiers for automatic detection of specification tables on a product webpage. %This simple layout representation of tables also make it more likely to only contain specification information on a product web page making it easier to identify the specification block amongst non-specification blocks. 
Unfortunately, several popular e-commerce websites (like, Reliance Digital, Scandid, Paytm Mall and others) make use of non-table and non-list tags for rendering specification blocks and therefore, the above approaches fail to generalize to them. %We are concerned with capturing information from specification blocks constructed out of HTML fragments not limited to tables and lists.  

In this paper, we are concerned with the problem of extracting product specifications (attribute-value pairs) from specification blocks independent of the specific HTML tags used for their rendering. Irrespective of the specific HTML tags used for their rendering, specification blocks are typically composed of repeating sub-blocks that are usually consistent in their usage of tag patterns. We leverage this spatial invariance and consistency of tag patterns to propose a scalable specification extraction solution. Our solution comprises two components\textemdash (a) specification block classification; and (b) attribute-value extraction. Using a combination of hand-coded features and deep learned spatial and token features, we propose a two-stage cascade block classification model that identifies specification blocks among all the HTML blocks present on a product webpage. Our extraction model is based on typical wrapper induction techniques~\cite{dalvi2011automatic,kushmerick1997wrapper} that use a set of attributes to learn wrappers and the induced wrappers to extract more attributes, in an iterative fashion, using a seed attributes list to bootstrap the process. Starting from the raw HTML text of a product webpage as input, we first use our block classification model to identify the product specification blocks on the page and then use our extraction model to extract the attribute-value pairs from these blocks. %We devise a two-step cascaded classification process to identify specification blocks in a product web page and use our extraction algorithm to capture their corresponding attribute-value pairs. Fig.\ref{fig:inspect_elements_example} shows two examples of product specifications that have diverse HTML fragments which our approach can successfully classify and extract.

We make these key contributions in this paper\textemdash (a) we propose a scalable product specification extraction approach that generalizes to the wide variety of specification blocks across product websites. To the best of our knowledge, none of the previous works have focused on extracting specification information from blocks beyond tables and lists. (b) we create a large scale labeled data of product specifications extracted from 14,111 diverse blocks taken from a range of different product websites. We make this data publicly available \footnote{https://drive.google.com/drive/u/4/folders/1vNkLbkqsaslxyXZuomzwmNZ7oXIj0s7V}. (c) we conduct thorough evaluation of our model, the components therein and show that the results corroborate with our claim.

\section{Related Work}
A body of work on Web tables~\cite{cafarella2008uncovering,cafarella2008webtables,limaye2010annotating} leverages HTML tables to extract relational data from the Web and linking the entity and relation mentions to a structured catalog. In their system, WebTables~\cite{cafarella2008uncovering,cafarella2008webtables}, Cafarella {\em et al.} present techniques for high-quality relation recovery from HTML tables on the Web. Using techniques involving hand crafted parsers and machine learning models, they create a large scale relations corpus of 154 million tables. They then present relation ranking approaches for searching within this corpus of tables. Deep learning-based approaches~\cite{10.5555/3298239.3298265} have also been applied to the table classification task. In their system, TabNet, they propose a DNN architecture based on RNN and CNN to capture the semantic structures of Web tables.

Specifically, for tables containing product specifications, Qiu et al.~\cite{qiu2015dexter} present DEXTER, a system to discover product sites and detect and extract product specifications from them. Unlike the work on web tables, their supervised learning-based approach extends to HTML fragments (tables and lists) present on web pages and classifies them as product specifications or not. For extraction of attribute-value pairs from the product specifications, they propose two approaches - (1) an unsupervised approach, inspired from other works like RoadRunner~\cite{crescenzi2001roadrunner}, that leverages the homogeneity in the structure of product specifications across web pages; (2) their other approach is inspired from that by Dalvi {\em et al.}~\cite{dalvi2011automatic} and involves inference of extraction patterns based on annotations generated by automatic but noisy annotators. Petrovski {\em et al.}'s specification extraction work~\cite{10.1145/3106426.3106449} builds upon DEXTER. In addition to introducing additional features for specification detection, they also propose a classification approach to identify attribute and value columns, thereby allowing for specifications extraction from tables with more than two columns.  

We also draw motivation from work on wrappers and wrapper induction~\cite{dalvi2011automatic,kushmerick1997wrapper} that deal with the problem of inducing wrappers from labeled examples. Here, wrappers are rules, typically based on xpath or DOM paths, used for information extraction from web pages.

A more recent work in this area uses visual input of tables instead of textual HTML data ~\cite{DBLP:journals/corr/abs-2103-05110}. Their method relies on taking rendered images of web tables and using CNN based VGG16 and ResNet50 with transfer learning to distinguish between genuine tables and layout tables. Again, the task here is focused on only classification of tables and also does not involve table information extraction. Computer vision is again applied by Gundimeda {\em et al.} who have developed a vision based approach that first evaluates the quality of food product images and then detects and extracts product attributes such as brand name, title, net weight, nurtritional facts using Optical Character Recognition~\cite{10.1007/978-981-13-1580-0_20}. This system therefore relies on the availability of product images and works only on products belonging to food category.

Further work on attribute-value extraction is formulated by Wang {\em et al.} where they rely on a question-answering model as a tool to extract values for  corresponding attributes~\cite{10.1145/3394486.3403047}. Here each attribute is considered as a question and a context such as title or description of the corresponding product is provided from which the value is determined as answer. This is helpful in scenarios with missing values and can use the context to extract them if present. 

\section{Our Approach}
\begin{figure}[htp]
    \centering
    \includegraphics[width=\columnwidth]{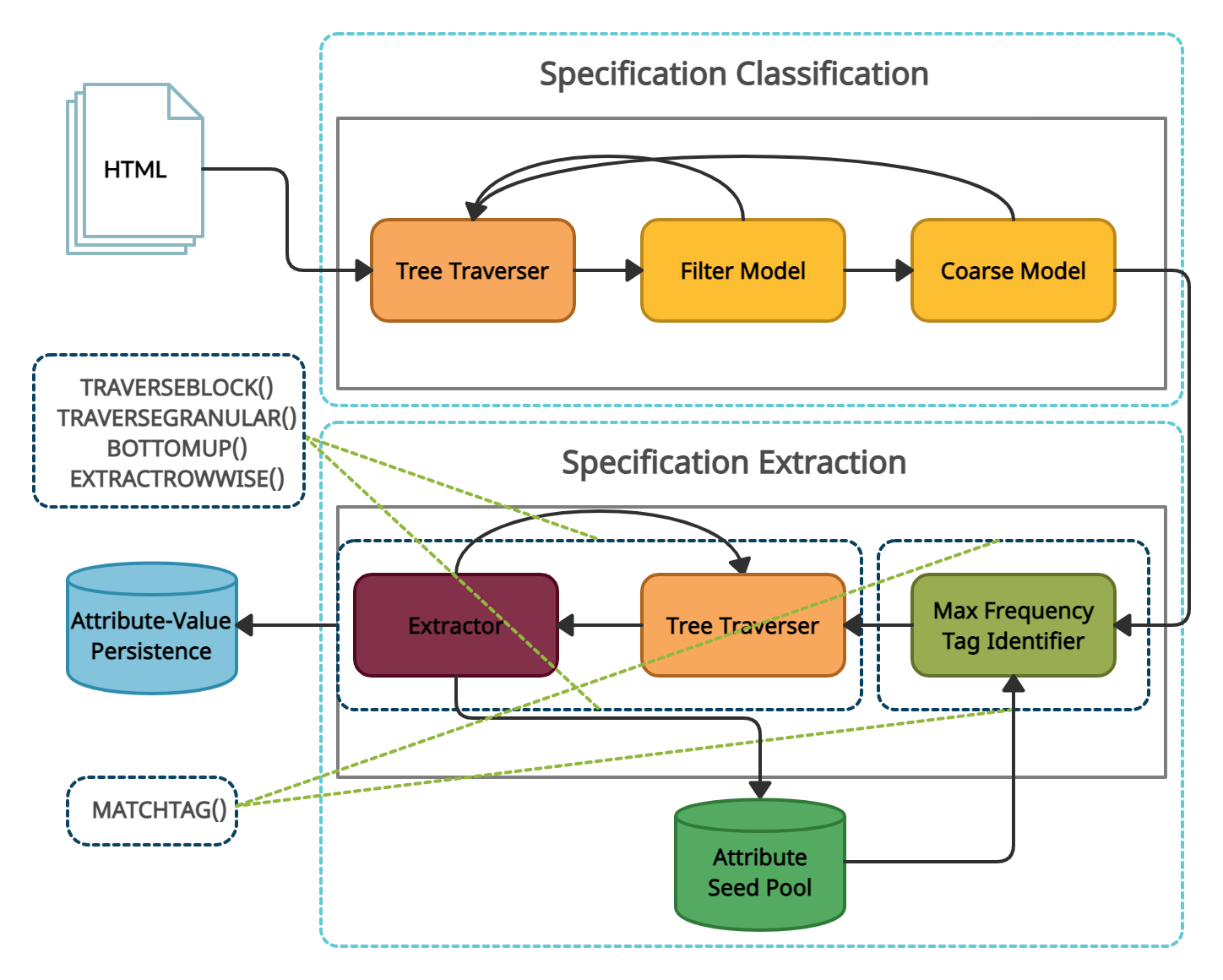}
    \caption{System Architecture}
    \label{fig:architectureMain}
\end{figure}

We propose a two-stage approach comprising a \textit{specification classification} stage followed by a \textit{specification extraction} stage. Figure \ref{fig:architectureMain} illustrates our overall solution.  %We start by taking the raw HTML of a product web page as input. %The overall end-to-end framework, as shown in Figure \ref{fig:architectureMain}, is divided into two stages\textemdash Specification Classification and Specification Extraction. The architecture also mentions pointers for the extraction algorithm's functions'(details in the extraction section) placements based on their order and step of execution. 
Starting with the raw HTML of a product webpage as input, in the specification classification stage, we are concerned with the problem of identifying blocks on the page that contain product specifications. We leverage the hierarchical DOM tree structure of the page and subject every DOM block through a binary classification decision of either containing product specifications or not. %Under the classification stage, we rely on the organization of the HTML's tree-based structure which provides the ability to traverse the DOM tree and localize and identify the node(s) containing relevant information, which in our case is the specification attributes and values. We employ two classification models, in cascade fashion, that accept these HTML blocks (nodes) as input and classify them as specification or non-specification blocks.
\begin{figure}[htp]
    \centering
    \includegraphics[width=\columnwidth]{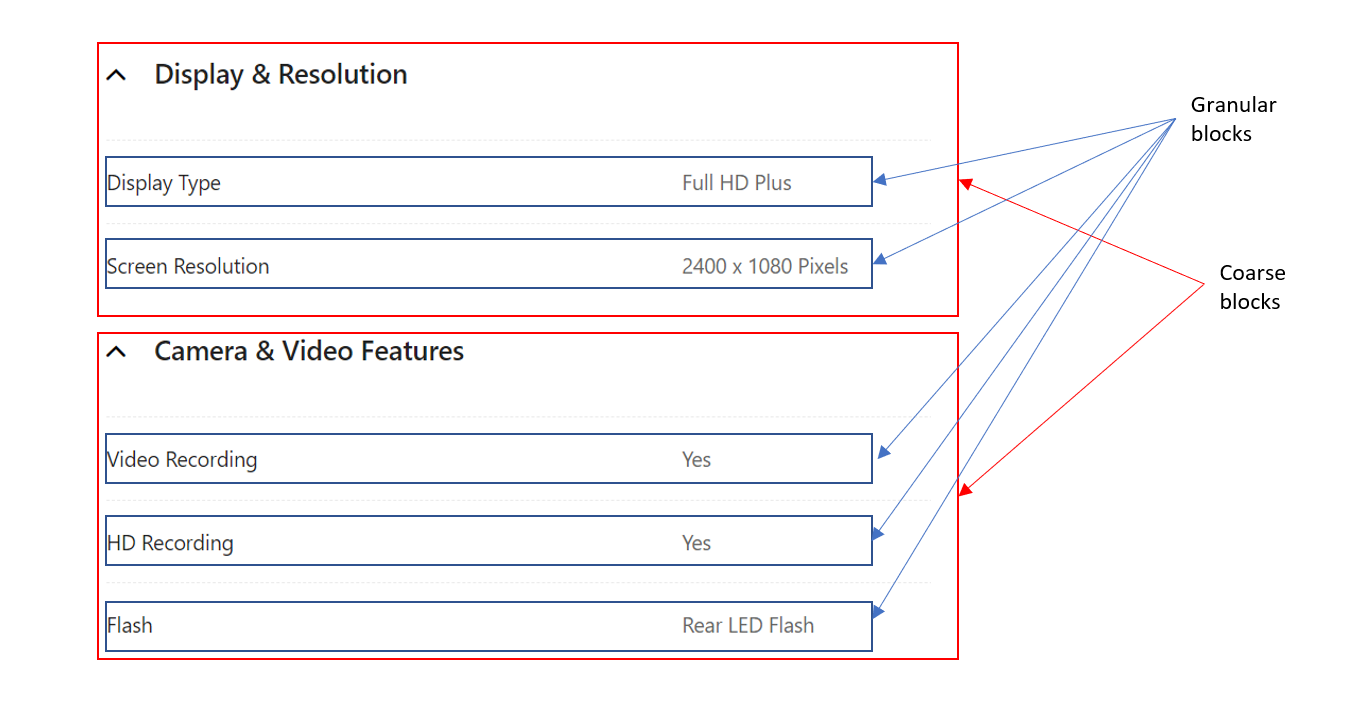}
    \caption{Coarse vs Granular blocks}
    \label{fig:coarse_granular}
\end{figure}
The output is a set of candidate specification blocks that are then passed on to the specification extraction stage. Our extraction model takes inspiration from wrapper induction techniques and leverages repeating wrapper patterns in the specification blocks to extract <attribute, value> tuples from the specification block candidates.  %Subsequently, in the specification extraction stage, we will be again taking advantage of this tree-based structure to extract the specification attribute-value pairs from the earlier candidate blocks with the introduction of an extraction method. The final output is a set of attribute-value tuples of the form <attribute, value>. 
We also interchangeably refer to a block immediately surrounding an <attribute, value> tuple as the ``granular'' block and the immediate block containing multiple such granular blocks as the ``coarse'' specifications block (Refer to Figure~\ref{fig:coarse_granular}). Thus, the responsibility of the specification classification stage is to identify coarse specification blocks and that of the extraction stage is to extract the specification tuples from the granular blocks contained within the coarse blocks.

%We have defined the terms ``coarse'' and ``granular'' to indicate the two kinds of structures within an HTML block that will be used in the following sections. Figure \ref{fig:coarse_granular} shows two specification blocks which are also coarse blocks and five granular blocks. A coarse block consists of multiple children nodes (typically more than two). In terms of specifications, each coarse block is a tabular specification block whereas for non-specifications, they are blocks that contain more than two children. Rows, cells and individual text fields come under granular blocks and thus each coarse block will contain at least one granular block.

In the following sections, we describe our specification classification and extraction models in more detail.

\subsection{Specification Classification}

\begin{algorithm}[htp]
\DontPrintSemicolon
\SetKwFunction{sTraverse}{specTraverse}%
\SetKwProg{Fn}{Function}{:}{}
\Fn{\sTraverse{$node, candidate$}}{
    \If{$node.name \notin blackList$ \textit{and} $node \supseteq Text$}{
        \If{$node.Children.length > 1 $}{
            \If{$filterModel(node) \in specification $}{
                \If{$coarseModel(node) \in specification $}{
                     $candidate \gets candidate\cup node$ \;
                     $node.decompose()$ \;
                }
                
            }
        }
        \For{$block \in node.Children$}{
              \sTraverse{$block, candidate$} \;
        }
    }
}

\caption{Specification Classification\label{alg:ALG_1}}
\end{algorithm}
Given a HTML DOM block $\mathbf{x}$, we would like to classify it as $y=f(\mathbf{x})$, where, $y\in\{1, 0\}$ depending on whether or not the node contains product specifications and $f(\cdot)$ is a binary classifier. As classifier, we use a cascade model comprising %In order to predict whether a given HTML DOM node is a specification block or not, we use 
a Support Vector Machine (SVM) followed by a deep Convolutional Neural Network (CNN) with word embeddings. %The purpose of using a light-weight SVM model is to filter out nodes that don't contain specification information. 
A product web page typically consists of a large number of blocks and using a deep CNN model for prediction on each of these incurs high inference cost. The computationally efficient SVM model serves as a ``filter'' model leaving a relatively smaller number of blocks to be further classified by the coarse CNN model. Starting with the HTML <body>, we recursively traverse the DOM blocks on the page, subjecting each through our cascade classification model, leading to a set of candidate specification blocks as the output from this stage. %Tree traversal on an HTML page is performed starting with <body> of the tree and visiting all of its children nodes. Each of these nodes may further contain more children and this process of tree traversal continues recursively with classification being done by the filter and coarse models at each node to ascertain whether they are specification block or not, or till it reaches the leaf nodes and stops. In case of a specification block, that node along with all of its children and lower-level nodes/branches are removed in place from the DOM tree to be set aside as a potential candidate block for the extraction module. 
We describe our specification block classification approach in Algorithm~\ref{alg:ALG_1}.% shows the tree traversal sequence in which the filter and coarse models' classifications are applied while traversing through each HTML DOM node.

%testing out algorithmic package function 

\subsubsection{Filter Model}
The existing specification classification approaches, based on tables and lists, derive their features from the structure of the blocks ({\em e.g.} number of columns, number of list elements {\em etc}). As features for our SVM-based filter model 
%Due to the nature of the HTML blocks that don't necessarily contain tables or lists, it introduces difficulty when it comes to identifying features that could work on all the HTML structures not limiting to tables or lists. Keeping this in mind, 
we have engineered the following features that are generic to all types of HTML structures beyond tables and lists: total number of text fields, total text length, alpha-numeric ratio, number of images, number of links, and uppercase-total text length ratio.

\subsubsection{Coarse Model}
\begin{figure}[htp]
    \centering
    \includegraphics[width=8.5cm]{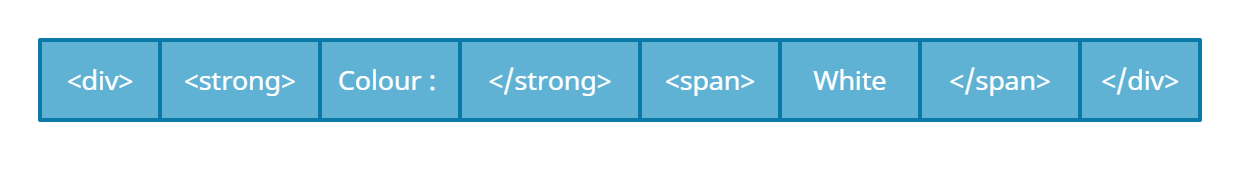}
    \caption{Tokenized HTML}
    \label{fig:word_tokens}
\end{figure}
\setcounter{tocdepth}{2}
\setcounter{secnumdepth}{4}

\begin{figure}[htp]
    \centering
    \includegraphics[width=8.5cm]{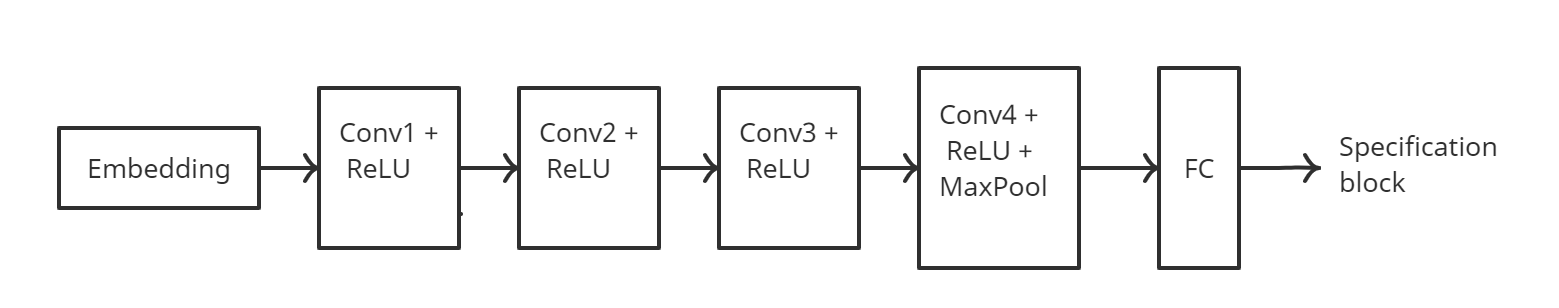}
    \caption{Word CNN Architecture}
    \label{fig:word_cnn_architecture}
\end{figure}

The specification blocks identified by the filter model are first tokenized (Refer to Figure~\ref{fig:word_tokens} for a sample) using a tokenizer that removes any references to HTML class or id attribute details. % as they vary from source to source, and 
The tokenized block text is then encoded using a pre-trained word embedding model into a fixed dimension vector~\cite{mikolov2013efficient, rehurek_lrec} and passed as input to the CNN model as shown in Figure~\ref{fig:word_cnn_architecture}. The output of the model is again a binary label indicating whether the input block is a specification block or not.

\subsection{Specification Extraction}

\begin{algorithm}
\setcounter{AlgoLine}{0}
\caption{Extraction\label{alg:ALG_2}}
\DontPrintSemicolon
\SetKwFunction{mTag}{MatchTag}%
\SetKwProg{Fn}{Function}{:}{}

%\Require T: an HTML tree structure, S: a set of attribute names, B: a set of blacklisted tags
%\Ensure M: an empty set
\Fn{\mTag{$T, M$}}{
    \If {$T.name \notin blackList $}{  
        \If {$T.type \neq Text $}{
            \For{$ \mbox{each node n} \in T.children$}{
                \If {$n.type = Text$}{
                    \If {$n \in S$}{
                        $M \gets M\cup n.parent $ \;
                    }
                }
                \mTag{$n, M$}
            }
        }
    }

    \Return{$M$}
}
\SetKwFunction{eRowWise}{ExtractRowWise}%
\SetKwProg{Fn}{Function}{:}{}
\Fn{\eRowWise {e}}{
    $counter \gets 0$\;
    $A \gets \phi$\;
    $V \gets \phi$\;
    \For {$\mbox{each item i} \in e.descendants$}{
        \If {$i.type = Text$ \textit{and} $ i.length > 0 $ \textit{and} $ i \notin X $}{
            $counter \gets counter + 1 $\;
        }
    }
    \If {$counter > 1$}{
        $posCounter \gets 0$ \;
        \For {$\mbox{each item i} \in e.descendants$}{
            \If {$i.type = Text$ \textit{and} $ i.length > 0$ \textit{and} $ i \notin X $}{
                \If{$counter = 2$}{
                    \If {$posCounter \bmod 2 = 0$}{
                        $posCounter \gets posCounter + 1$ \;
                        $A \gets i$\;
                    }
                    \Else{
                         $posCounter \gets 0$\;
                         $V \gets i$\;
                    }
                    
                }  
            }
        }
    }
\Return{$A, V$}
}

\end{algorithm}
%\floatstyle{nocaptionruled}
%\restylefloat{algorithm}
\RestyleAlgo{tworuled}
\begin{algorithm}[]
\DontPrintSemicolon
\SetKwFunction{bUp}{BottomUp}%
\SetKwProg{Fn}{Function}{:}{}
\Fn{\bUp{e}}{
    \If {$e.parent$}{
        $counter \gets 0$ \;
        \If {$e.parent.type \neq Text $}{
            \For {$\mbox{each item i} \in e.parent.descendants$}{
                \If {$i.type = Text$ \textit{and} $i \notin X $}{
                    $counter \gets counter + 1 $ \;
                }
            }
            \If {$counter > 1$}{
                $P \gets P \cup {\eRowWise{e.parent}}$ \;
            }
        }
        \Else{
            {\bUp{e.parent}}
        }
    }
    \Else{
       {\bUp{e.parent}} 
    }
    
\Return{$P$}
}

\SetKwFunction{tGranular}{TraverseGranular}%
\SetKwProg{Fn}{Function}{:}{}
\Fn{\tGranular{T, M}}{
    $AV \gets \phi$ \;
    \For{$ \mbox{each element e} \in T.descendants$}{
        \If {$e.type = Text$ \textit{and} $e.parent.name = M$}{
            $AV, node \gets {\bUp{e, T}}$ \;
        }
    }
    \If{$node$}{
        \For{$sibling \in node.siblings$}{
            $AV \gets AV \cup {\eRowWise{sibling}}$ \;
        }
    }
\Return{$AV$}
}
\SetKwFunction{tBlock}{TraverseBlock}%
\SetKwProg{Fn}{Function}{:}{}
\Fn{\tBlock{T, M, AV}}{
    \If {$T.getText$ \textit{and} $T.name = M$}{
        $AV \gets AV \cup {\tGranular{T, M}}$ \;
    }
    \Else{
        \For{$ \mbox{each node n} \in T.children$}{
            {\tBlock{$n, M, AV$}} \;
        }
    }
   \Return{$AV$}
}
\end{algorithm}

Our specification extraction approach takes inspiration from wrapper induction-based techniques. For every candidate coarse specification block identified in the previous stage, we use known attribute names/values in its granular blocks to induce HTML wrappers and use the wrappers to extract further <attribute, value> pairs. We use a small set (Denoted by $S$ in Algorithm~\ref{alg:ALG_2}) of attribute names/values to bootstrap the process. %We are again leveraging the DOM tree structure to capture the specification attributes and their corresponding values using the methods defined in Algorithm \ref{alg:ALG_2}. During this stage, we also make sure to handle cases where- either the classification models fail to correctly classify a specification block and instead output non-specification blocks, or when the models classify a higher level block which also includes the specification blocks but is also surrounded by non-specification blocks. We use a bootstrap 
Below we briefly describe the different functions that are part of the extraction algorithm:
%\subsubsection{\ref{alg:ALG_2}\ref{}}

%{\mTag}: From every candidate specification block $T$, we look for its granular blocks containing mentions of attributes from the seed attribute set $S$. We then extract HTML wrappers from all such mentions and pick the wrapper $M$ with the maximum \textit{support}. Here, support of a wrapper is defined as the number of other attributes extracted, with the help of this wrapper, from the block $T$. This simple heuristic works well in filtering out any spurious noisy wrappers  
%Plays a crucial role in recognizing the HTML tags associated with attributes in the candidate blocks with that of the attributes in the seed pool. Only identifying the tag can result in the wrong HTML block being identified, especially when both the specification and non-specification blocks are constructed using the same HTML tag(s) such as div. This scenario is more evident in the cases where the classification models generate candidate blocks that have specification blocks surrounded by multiple levels of non-specification blocks. This leads to similar tags from non-specification blocks being incorrectly considered after matching attribute(s) for extraction. To minimize this error, we identify the attributes in the blocks that exist in the seed pool. Once identified, we calculate the frequency of these attributes' tags. 
 {\mTag}: From every candidate block $T$, we look for its granular blocks containing mentions of attributes from the seed attribute set $S$. We then extract HTML wrappers from all such mentions and pick the wrapper $M$ with the maximum \textit{support}. Here, support of a wrapper is defined as the number of attributes present in both the block $T$ and the seed pool, for that wrapper. This simple heuristic works well in filtering out any spurious noisy wrappers. The wrapper with the maximum support along with its support attributes is sent as input to {\tBlock}. This keeping a record of the support attributes in addition to their corresponding wrapper comes in handy in localizing the actual specification block, especially in cases where both the specification and non-specification blocks are constructed using the same wrapper.
%we calculate the frequency of attributes in each candidate block that already exists in our seed pool, and only the tag with maximum frequency qualifies. Along with this tag, its respective attributes present in the candidate block that already existed in the seed pool are sent as input to {\tBlock}.

{\tBlock}: A top-down tree traversal is performed here and the granular HTML block (text cell containing attribute) that is associated with the wrapper and the matched attribute name identified by {\mTag} is selected. Simply identifying the text cell's wrapper neither fetches all the attributes and values present in that candidate block nor helps in locating this granular block amongst the entire candidate block. For this we introduce {\tGranular} and {\bUp}.

{\tGranular}: This performs the task of first finding the actual row containing an attribute-value pair with the granular block provided by {\tBlock}. This requires a bottom-up traversal performed by {\bUp}. Once this row is identified, the next step is to find all the sibling rows, each containing an attribute-value pair. As preferred, specification block titles get ignored in cases where it is not composed of pair-wise text fields.

{\bUp}: Repeated bottom-up traversal is performed from the column granular block output by {\tBlock} until at least two text fields are identified to represent attribute-value pair. Once ascertained, {\eRowWise} is called to extract the corresponding attribute-value pair.

{\eRowWise} is configurable with respect to the number of \linebreak columns\textemdash be it a row containing either 2 or 4 columns, amounting to 1 or 2 attribute-value pairs per row.

\section{Curation of the Product Spec Block Dataset}
Existing product specification datasets are limited to blocks composed of HTML tables and lists. Since our approach extends to blocks beyond HTML tables and lists, we curate a custom dataset comprising of blocks composed of a range of different HTML tags. Our dataset represents the %We have created custom datasets that consist of HTMLs with tags other than tables and lists inside the specification blocks to demonstrate the complete functionality of our solution on a 
diverse HTML structures typically used to present product specifications across product categories on different product websites. %Our datasets have been distinguished for two purposes- Classification and Extraction. For both of these, we prepare them by making use of the same collection of HTMLs from six different websites/sources using ~\cite{richardson2007beautiful}. 
Table~\ref{tab:rawHTMLData} illustrates how the entire dataset has been divided into three types- Train, Validation, and Holdout. Each pair of Train and Validation datasets are sourced from categories belonging to one website each, while the Holdout set comprises of HTMLs from three new websites. The website names along with their respective categories to which the HTMLs belong to are indicated in Table~\ref{tab:datasetDescription}. 

% \begin{table*}[htp]
%     \centering
%     \caption{Caption}
%     \label{tab:my_label}
%     \begin{tabular}{*{2}cp{15mm}p{25mm}r>{\raggedright}p{15mm}*{3}{>{\raggedleft}p{10mm}}>{\raggedleft\arraybackslash}p{10mm}}
%          Dataset & Type & Source & Category & \#Webpages & Spec tags & Spec blocks & Non-spec blocks & Attr-Val pairs & Unique attributes  \\ \hline
%          TD-1 & Train & \multirow{2}{15mm}{Reliance Digital} & \multirow{2}{25mm}{Laptop, Refrigerator, AC, Washing Machine, TV} & 767 & \multirow{2}{15mm}{ul, div, li, a, br} & 4,804 & 20,469 & &  \\
%          VD-1 & Validation & & & 193 & & 1,546 & 6,070 & 8,098 & 188 \\
%          TD-2 & Train & \multirow{2}{15mm}{Scandid} & \multirow{2}{25mm}{Laptop, TV, Smartphone} & 537 & \multirow{2}{15mm}{div, span, h2, strong} & 425 & 12,982 & &
%     \end{tabular}
% \end{table*}

\begin{table}[h]
    \centering
    \caption{Raw HTML Dataset}
    \begin{tabular}{|c|c|r|}
    \hline
      Dataset & Type & \#HTML \\
      \hline
      TD-1 &  Train & 767\\
     
      VD-1 & Validation & 193 \\
      
      TD-2 & Train  & 537\\
      
      VD-2 & Validation &  115 \\
      
      TD-3 & Train & 1,649 \\
      
      VD-3 & Validation &  413 \\
      
      HD-1 & Holdout & 2,237 \\
      
      HD-2 & Holdout & 157 \\
      
      HD-3 & Holdout & 131 \\
   
      \hline
      
    \end{tabular}
    
    \label{tab:rawHTMLData}
\end{table}

\subsection{Classification Dataset}

\begin{table}[htp]
    \centering
    \caption{Dataset Description}
    \begin{tabular}{|c|c|c|}
    \hline
     Dataset & Source & Category \\
    \hline
    TD-1 &  \multirowcell{2}{Reliance \\ Digital} & \multirowcell{2}{Laptop, Refrigerator, \\ AC, Washing Machine, TV} \\
    \cline{1-1}
    VD-1 & &  \\
    \hline
    TD-2 &  \multirowcell{2}{Scandid} & \multirowcell{2}{Laptop, TV, \\ Smartphone} \\
    \cline{1-1}
    VD-2 & &  \\
    \hline
    TD-3 &  \multirowcell{2}{CDHFineChemical} & \multirowcell{2}{Chemical} \\
    \cline{1-1}
    VD-3 & & \\
    \hline
     \multirowcell{2}{HD-1} & \multirowcell{2}{Paytmmall} & \multirowcell{2}{Laptop, TV, \\ Smartphone} \\
     & &  \\
     \hline
      HD-2 & Somanyceramics & Tiles \\
      \hline
      \multirowcell{2}{HD-3} & \multirowcell{2}{Alibaba} & \multirowcell{2}{Animal Feed, Dairy, \\ Beddings, Skin Care} \\
      & &  \\
      \hline
    
    \end{tabular}
    
    \label{tab:datasetDescription}
\end{table}

For classification, we have utilized the inherent tree-based structure to generate the coarse blocks. In order to extract non specification blocks, we traverse the tree visiting each HTML node and identifying whether the given node has more than one child and whether that node contains text at any depth. The specification blocks have been removed in advance to prevent mistakenly tagging them as non-specification. We have ignored the first few blocks that we extracted as part of non-specification to avoid the blocks that are on the top most level of the tree. Since there can be more than one specification block in some of the product pages, we have considered them as separate specification blocks. Even after considering multiple specification blocks per product page, there is a large imbalance between the number of specification and non-specification blocks which is inherent to product web pages. This imbalance is visible in Table \ref{tab:clfnData}. Table \ref{tab:clfnData} also shows the different HTML tags present in the specification block of each dataset's web page.

\begin{table}[htp]
    \centering
    \caption{Classification Dataset}
    \begin{tabular}{|c|c|c|c|}
    \hline
      Dataset &  \#Spec & \#Non-spec & Spec Tags \\
      \hline
      TD-1 & 4,804 & 20,469 & \multirowcell{2}{ul, div, \\ li, a, br} \\
      \cline{1-3}
      VD-1 & 1,546 & 6,070 & \\
      \hline
      TD-2 & 425 & 12,982 & \multirowcell{2}{div, span, \\ h2, strong} \\
      \cline{1-3}
      VD-2 & 115 & 3,197 & \\
      \hline
      TD-3 & 1,649 & 10,713 & \multirowcell{2}{div, p, a} \\
      \cline{1-3}
      VD-3 & 413 & 2,799 & \\
      \hline
      \multirowcell{2}{HD-1} & \multirowcell{2}{4,871} & \multirowcell{2}{35,704} & \multirowcell{2}{div, span, i, h2,  \\ br, a, b, font, img} \\
      & & & \\
      \hline
      \multirowcell{2}{HD-2} & \multirowcell{2}{157} & \multirowcell{2}{6,710} & \multirowcell{2}{div, strong, \\ span} \\
      & & & \\
      \hline
      \multirowcell{2}{HD-3} & \multirowcell{2}{131} & \multirowcell{2}{11,918} & \multirowcell{2}{div, span, \\ dl, dt, dd} \\
      & & & \\
      \hline
    \end{tabular}
    
    \label{tab:clfnData}
\end{table}

\subsection{Extraction Dataset}
Since we aren't training any model for our extraction procedure, we'll be using only the validation and holdout datasets containing HTML with their corresponding page-wise attribute-value pairs. For an attribute-value pair to be considered valid, we include only those attributes whose values are not empty. We have also come across repetition of the attribute-value pair in the same spec block and we consider only a single occurrence of them. The extraction dataset details are shown in Table \ref{tab:my_label}.

\begin{table}[htp]
    \centering
    \caption{Extraction Dataset}
    \begin{tabular}{|c|c|c|}
    \hline
        \multirowcell{2}{Dataset} & \vtop{\hbox{\strut \hfil \#Attr-Val}\hbox{\strut Pairs}} & \vtop{\hbox{\strut \hfil \# Unique}\hbox{\strut Attributes}}\\
        \hline
        VD-1 & 8,098 & 188 \\
        %\hline
        VD-2 & 5,869 & 382 \\
        % \hline
        VD-3 & 3,016 & 8 \\
        %\hline
        HD-1 & 51,200 & 533 \\
        %\hline
        HD-2 & 1,546 & 10 \\
        %\hline
        HD-3 & 1,801 & 205 \\
        \hline
    \end{tabular}
    
    \label{tab:my_label}
\end{table}

\section{Evaluation}
\subsection{Experiment Setup}
\subsubsection{Model Configuration}
An SVM with linear kernel is used as the filter model. The TD-* datasets are used for training the filter model. For the coarse model, we use a word embedding of 100 dimensions that was pre-trained on the training and holdout sets presented in Table \ref{tab:clfnData}. This input is passed through a dropout layer to a 4-layer CNN with ReLU activation after each convolutional layer. Max pooling is only performed after the fourth and final ReLU followed by another dropout layer in to a linear fully connected layer. Batch size used is 2, number of filters is 24 for each convolutional layer with a size of 4 and dropout is set to 0.4. An Adam optimizer~\cite{kingma2017adam} is employed with a learning rate of $1\times10^{-5}$ and L2 regularization  $1\times10^{-6} $%\num{1e-6}. 

The training set for the coarse model comprises of 19,940 unique samples across the three sources with an input length of 40 word tokens with digits and punctuation removed except for <, > and '/'.

\subsubsection{Extraction Configuration}
We started with an initial seed pool of 45 attributes and added a feedback loop from the extraction algorithm to the seed pool for real-time enrichment of the extracted attributes. We consider an extracted attribute value pair to be true positive only if both the attribute and the value match one-on-one with the ground truth attribute-value pair belonging to its corresponding product web page.

\subsection{Results}
\subsubsection{Classification}
For baseline performance comparison, we compared our method with Petrovski {\em et al.}~\cite{10.1145/3106426.3106449} which supports only tables and lists. Due to this limitation, we created a new dataset containing specifications and non-specifications constructed with tables and performed a 5-fold cross validation, with the performance results from an additional test set as the final comparison. Due to the low quantum of table data samples, we could only validate it on our filter model. Also since our approach hasn't been made suitable to use any delimiters unlike ~\cite{10.1145/3106426.3106449}, we didn't go ahead with conducting a comparison on list data. We have also conducted the classification experiments on our validation and holdout datasets as seen in Table \ref{tab:filterAndCoarseM}.

\begin{table}[htp]
    \centering
    \caption{Specification and non-specification detection for table blocks}
    \begin{tabular}{|c|c|c|c|}
    \hline
    Approach & Precision & Recall & F1-Score \\
    \hline
    Petrovski {\em et al.} \cite{10.1145/3106426.3106449} & 0.936 & 0.935 & 0.936 \\
    \hline
    Filter SVM (Our) & 0.951 & 0.949 & 0.950 \\
    \hline
    \end{tabular}
    
    \label{tab:comparisonTableClassification}
\end{table}

\begin{table}[htp]
    \centering
    \caption{Filter and Coarse model classification performance w.r.t. specification blocks}
    \begin{tabular}{|c|c|c|c|c|c|c|}
    %heading
    \hline
    \multirowcell{2}{Dataset} & \multicolumn{3}{c|}{Filter SVM}  & \multicolumn{3}{c|}{Coarse CNN}\\
    \cline{2-7}
    \centering
     &  P & R & F1  & P & R & F1 \\

    \hline
    VD-1 & 0.412 &0.994 & 0.583 & 0.921 & 0.993 & 0.955\\
    \hline
    VD-2 &  0.184 & 1.0 & 0.311 & 0.905 & 0.991 & 0.946\\
    \hline
    VD-3  & 0.998 & 1.0 & 0.999 & 1.0 & 1.0 & 1.0\\
   \hline

   HD-1 & 0.179 & 0.978 & 0.303 & 0.548 & 0.995 & 0.707\\
   \hline
   HD-2 & 0.073 & 1.0 & 0.136 & 0.149 & 1.0 & 0.260\\
   \hline
   HD-3 & 0.026 & 1.0 & 0.051 & 0.056 & 1.0 & 0.107\\
   \hline
   
    \end{tabular}
    
    \label{tab:filterAndCoarseM}
\end{table}

\subsubsection{Extraction}
We have again compared our extraction performance with that of ~\cite{10.1145/3106426.3106449}. We sampled columns of specification tables from the previous dataset and annotated them as attributes and values. A 5-fold cross validation is performed. Since our extraction approach does not require training, we have used a common test set split for both the approaches' evaluations.
\begin{table}[htp]
    \centering
    \caption{Attribute-value extraction for table specification blocks}
    \begin{tabular}{|c|c|c|c|}
    \hline
    Approach & Precision & Recall & F1-Score \\
    \hline
    Petrovski {\em et al.} \cite{10.1145/3106426.3106449} & 0.976 & 0.976 & 0.976 \\
    \hline
    Extraction (Our) & 0.962 & 0.861 & 0.909 \\
    \hline
    \end{tabular}
    
    \label{tab:comparison_table_extraction}
\end{table}

\subsubsection{End-to-End}
For the end-to-end performance of classification and extraction, we have considered three arrangements of classification: Filter SVM, Coarse CNN and Filter SVM + Coarse CNN. The candidate blocks classified as specification from each of these cases are given as input to the extraction module. The results are presented in Tables \ref{tab:FilterendToEnd},  \ref{tab:CoarseendToEnd} and \ref{tab:endToEnd}.

\begin{table}
    \centering
     \caption{End-to-end performance Filter+Extraction}
    \begin{tabular}{|c|c|c|c|c|}
    \hline
    
   Dataset & Precision & Recall &F1-Score \\
    \hline
     VD-1  & 0.790 & 0.977 & 0.874  \\ 
    \hline
     VD-2  & 0.894 & 1.0 & 0.944  \\ 
     \hline
     VD-3 & 1.0 & 0.995 & 0.997  \\
     \hline
     
    HD-1 &  0.639 & 0.932 & 0.758   \\
    \hline
    HD-2 &  0.782 & 1.0 & 0.878  \\  
    \hline
    HD-3 &  0.327 & 0.993 & 0.492  \\
     \hline
     
    \end{tabular}
   
    \label{tab:FilterendToEnd}
\end{table}

\begin{table}
    \centering
     \caption{End-to-end performance Coarse+Extraction}
    \begin{tabular}{|c|c|c|c|}
    \hline
    
    Dataset & Precision & Recall &F1-Score \\ % \vtop{\hbox{\strut \hfil Avg}\hbox{\strut Run Time (sec)}}\\
    \hline
     VD-1  & 0.822 & 0.965 & 0.888  \\ 
    \hline
     VD-2  & 0.688 & 1.0 & 0.815  \\ 
     \hline
     VD-3 & 1.0 & 1.0 & 1.0  \\
     \hline
     
    HD-1 &  0.685 & 0.994 & 0.811   \\
    \hline
    HD-2 &  0.907 & 1.0 & 0.951  \\  
    \hline
    HD-3 &  0.285 & 1.0 & 0.444   \\
     \hline
     
    \end{tabular}
   
    \label{tab:CoarseendToEnd}
\end{table}

\begin{table}
    \centering
    \caption{End-to-end performance Filter+Coarse+Extraction}
    \begin{tabular}{|c|c|c|c|}
    \hline
    
    Dataset & Precision & Recall &F1-Score \\%& \vtop{\hbox{\strut \hfil Avg}\hbox{\strut Run Time (sec)}}\\
    \hline
    VD-1  & 0.915 & 0.977 & 0.945  \\ 
    \hline
     VD-2  & 0.980 & 0.997 & 0.988  \\ 
     \hline
     VD-3 & 1.0 & 0.995 & 0.997  \\
     \hline
     
    HD-1 &  0.762 & 0.932 & 0.838   \\
    \hline
    HD-2 &  0.945 & 1.0 & 0.972  \\  
    \hline
    HD-3 &  0.547 & 0.997 & 0.706  \\
     \hline
     
    \end{tabular}
    
    \label{tab:endToEnd}
\end{table}

\begin{table*}[htp]
    \centering
    \caption{End-to-end specifics}
    \begin{tabular}{|c|c|c|c|c|c|c|c|c|c|}
    %heading
    \hline
    \multirowcell{2}{Dataset} & \multicolumn{3}{c|}{\# Avg Candidate Blocks}  & \multicolumn{3}{c|}{ Avg Classification Time (sec)} & \multicolumn{3}{c|}{Avg Extraction Time (sec)}\\
    \cline{2-10}
    \centering
     &  Filter & Coarse & Filter+Coarse  & Filter & Coarse & Filter+Coarse & Filter & Coarse & Filter+Coarse \\

    \hline
    VD-1 & 9.89 & 1.0 & 3.04 & 0.30 & 0.40 & 0.36 & 0.25 & 0.93 & 0.09\\
    \hline
    VD-2 &  20.56 & 1.78 & 1.97 & 0.58 & 1.60 & 0.63 & 0.35 & 39.5 & 0.23\\
    \hline
    VD-3  & 4.08 & 1.0 & 1.08 & 0.09 & 0.16 & 0.10 & 0.01 & 0.01 & 0.01 \\
   \hline

   HD-1 & 19.56 & 5.72 & 9.47 & 0.68 & 1.04 & 0.74 & 0.35 & 13.09 & 0.20\\
   \hline
   HD-2 & 7.33 & 4.94 & 4.36 & 1.78 & 1.62 & 2.29 & 0.51 & 0.49 & 0.44\\
   \hline
   HD-3 & 23.52 & 1.0 & 4.63 & 0.44 & 0.54 & 0.58 & 0.36 & 3.66 & 0.11\\
   \hline
   
    \end{tabular}
    
    \label{tab:specifics}
\end{table*}

\subsection{Observations}

The comparison results of table specification classification given in Table \ref{tab:comparisonTableClassification} shows that our model outperforms the baseline performance from ~\cite{10.1145/3106426.3106449} by 1.4\% in F1-Score, with the added flexibility of utilizing the same features across the non-table and non-list based specification blocks as well and thereby bringing in more coverage. For extraction, as visible from Table \ref{tab:comparison_table_extraction} for specification tables, our model falls behind the baseline performance from ~\cite{10.1145/3106426.3106449} by 6.7\%. This can be explained by the structuring of the columns in select rows where there are more than four text fields, including hyperlinks, even though the number of attribute-value pairs is only two. This led to incorrect combinations of attribute-value pairs getting captured, causing a comparatively high amount of false positives and false negatives. We can also see from Table \ref{tab:filterAndCoarseM} that both the filter and coarse model can maintain high recall, with the coarse model also having relatively high precision. The models have been seen to perform well w.r.t. recall, although precision has dropped on sources that haven't been trained on previously (HD-*) except for pre-trained embedding for the coarse model.

\begin{figure}[htp]
    \centering
    \includegraphics[width=8.5cm]{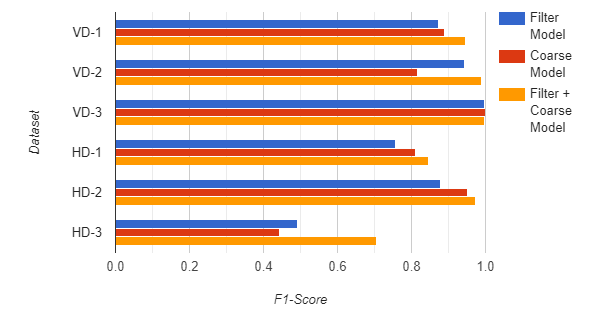}
    \caption{End-to-end Classification+Extraction F1 scores}
    \label{fig:End-to-endF1scores}
\end{figure}

The three different classification arrangements part of the end-to-end approach presented on Table \ref{tab:FilterendToEnd}, \ref{tab:CoarseendToEnd} and \ref{tab:endToEnd} and Figure \ref{fig:End-to-endF1scores} show that the two-step classification is crucial in achieving a high precision and coverage, and clearly outperforms the individual model configuration. We can also see that a higher precision of the coarse model did not always translate into a higher precision on its end-to-end arrangement. This varies between sources as to what amount of similar attribute names occur in the non-specification blocks. At the same time, a low classification precision of the filter SVM did not lead to a low precision during the end-to-end extraction. Still, in certain instances, it can happen at the expense of the extraction recall. A very high precision helps in narrowing down and localizing the specification block as close as to the true specification block leading to the extraction algorithm not coming across potential false positive attribute matches. This makes it less likely to extract from the wrong specification block amongst the candidate blocks in an ideal scenario such as VD-3.

Table \ref{tab:specifics} captures the number of candidate blocks generated and the individual run time of the classification and extraction on an average of the three end-to-end arrangements. Even though the expectation would be that the average classification duration would be much higher for the coarse model when compared to the filter model, it is also true for the extraction process on 4 out of 6 cases even when the number of candidate blocks for the coarse model is lower relative to the filter model. This is because the coarse model classifies a much higher level block (that contains the specification block as well) which leads to the extraction method trying to extract attribute-value pairs from a bigger block. Again, this varies from source to source based on the occurrence of attribute(s) outside the specification block, such as in the product description, and is also affected by the presence of text fields in a pair-wise format. Also, it can be noticed that the filter model does its job of reducing the time consumption during the classification stage as well as indirectly for extraction. 

As can be observed from Table \ref{tab:endToEnd}, our end-to-end approach is able to perform well on a diverse range of HTML structures covering different categories/segments and thus can generalize without explicitly tuning any of the component parameters. Our end-to-end two-step classification and extraction approach can maintain a low run time in achieving this outcome.

\subsubsection{\bfseries Does having a low precision affect the recall of the final extraction?}
Yes, it's entirely possible in scenarios where the models classify a bigger block (multiple levels higher than the specification block(s)) as specification which then contains a large amount of non-specification data in addition to specification data. The issue would surface during the incorrect (belonging to non-specification block) identification of the maximum wrapper support and the corresponding text due to a relatively higher quantum of matching seed attribute names belonging to non-specification which subsequently leads to an incorrect cell to be selected as the starting point of extraction. This results in actual specification wrappers getting excluded and thereafter the specification block(s). 

\subsubsection{\bfseries Does the filter model only assist in speeding up the classification process?}
Although, it would be expected that the run-time of a filter model based end-to-end extraction would be lower than filter+coarse model-based, that isn't the case as is evident from tables \ref{tab:FilterendToEnd}, \ref{tab:CoarseendToEnd} and \ref{tab:endToEnd}. This can be explained by the relatively lower precision of the filter model resulting in a higher number of candidate blocks and larger blocks which subsequently makes the extractor run on each one of them.
Even with a lower precision, the filter model is successful in rejecting a lot of non-specification blocks due to a considerable data imbalance between specification and non-specification, with the remaining false positives being handled by the coarse model. This two-step classification, in fact, helped the filter+coarse models to generate fewer candidate blocks but with high recall.

\subsubsection{\bfseries How is low recall related to an increase in classification and extraction time?}
A low recall can result in the classification models misclassifying the actual specification block as non-specification, which in turn makes the tree traversal method now visit each child of this specification block leading to an increase in the classification time considering that some specification blocks in our dataset contain as many as 100 rows. This could also shoot up the number of candidate blocks generated and thus causing the extraction time to go up as well.

\subsubsection{\bfseries Can low classification recall be offset by low classification precision for the end-to-end classification and extraction?}
Yes, it is possible. A lower recall means the actual specification block(s) is/are skipped. This is true for cases where an HTML contains either one or more than one specification block. However, in both the above cases, the low precision can cause a higher level block which ultimately contains either one, or in the latter case, multiple specification blocks to be wrongly classified as specification blocks, ultimately leading to all the specification blocks being present in the candidate block set as one. This will result in the final extraction to churn out most if not all of the attribute-value pairs that belong to these specification blocks.

\section{Conclusion and Future Work}
With an ever-increasing volume of product data being made available online, the use-cases requiring them, such as product search, fast and effective product comparison as well as recommendation engines, are growing manifold. But these applications require a concise form of product description that can only be obtained from the product specifications, which is a challenge as there is no strict adherence to limiting the usage to a standard set of HTML elements. In this paper, we have devised an approach that can identify and extract specification information in the form of attribute-value pairs over a wide range of frequently used HTML elements not limited to tables and lists. For the classification of specification block(s), we have used an SVM and a word-embedding-based CNN to handle the vast range of HTML elements defining the product web pages. To execute the extraction, we have developed an algorithm that can leverage the ordered tree structure of HTML to extract the attributes and their corresponding values. We have tested our approach on eleven categories across six websites. We also compared our approach on table specifications with that of ~\cite{10.1145/3106426.3106449} and have surpassed their table detection method by 1.6\% on F1-Score but registered a lower F1-score on the extraction task by a magnitude of 6.7\%. However, we have been able to demonstrate our method's ability to classify and extract attribute-value pairs from specification blocks composed of a multitude of HTML elements not restricted to only tables and lists.

The next step in this line of work is to add the capability to extract attribute-value pairs that use delimiters such as ":", "-" etc. Subsequently, an ontology of category-wise attributes can be created, which will open up more doors; for instance, ascertaining the most frequently occurring attributes in a particular category can assist in determining the attributes of a new product belonging to that category without performing any attribute extraction.

\bibliographystyle{ACM-Reference-Format}
\bibliography{spec}

\end{document}